\documentclass[runningheads]{llncs}

\usepackage{graphicx}

\usepackage{url}

\usepackage{breakurl}
\usepackage[colorlinks,allcolors={blue},breaklinks]{hyperref}

\usepackage{color}
\usepackage[dvipsnames]{xcolor}
\definecolor{newgrey}{HTML}{c2c2c2}

\usepackage{float}
\newfloat{mybox}{htbp}{lob}[section]
\floatname{mybox}{Box}

\usepackage{caption}

\usepackage{xurl}

\begin{document}

    \title{SHAPE: A Framework for Evaluating the Ethicality of Influence}

    \author{Elfia Bezou-Vrakatseli\inst{1}\orcidID{0009-0004-8954-246X} \and \\ Benedikt Br{\"u}ckner\inst{2}\orcidID{0000-0003-0699-1688} \and \\ Luke Thorburn\inst{1}\orcidID{0000-0003-4120-5056}}

    \authorrunning{E. Bezou-Vrakatseli et al.}
    
    \institute{
        King's College London, UK\\
        \email{\{elfia.bezou\_vrakatseli,luke.thorburn\}@kcl.ac.uk}\\[3mm] \and
        Imperial College London, UK\\ 
        \email{b.brueckner21@imperial.ac.uk}
    }

    \maketitle

    \begin{abstract}

        Agents often exert influence when interacting with humans and non-human agents. However, the ethical status of such influence is often unclear. In this paper, we present the SHAPE framework, which lists reasons why influence may be unethical. We draw on literature from descriptive and moral philosophy and connect it to machine learning to help guide ethical considerations when developing algorithms with potential influence. Lastly, we explore mechanisms for governing algorithmic systems that influence people, inspired by mechanisms used in journalism, human subject research, and advertising.

    
    \end{abstract}

\section{Introduction}

    Influence---which we define broadly as one agent taking an action that causes a change in another agent---is ubiquitous in multi-agent systems. If the agent being influenced is a person or is otherwise deserving of moral consideration, then it is widely accepted that some types of influence (e.g., blackmail, extortion) are unethical. In many settings where human communication is mediated by algorithms, however, the ethical status of influence is less clear. For example, interacting with a recommender system may change our preferences \cite{krueger2020,evans2022,carroll2021} and emotions \cite{kramer2014}, exposure to online political advertising can change our voting intentions \cite{coppock2020}, and interacting with large language models can change our opinions \cite{bai2023,jakesch2023}. In such cases, it can be easier to sense that there may be an ethical principle being violated than to articulate the principle of concern.

    There is a substantial body of work from descriptive and moral philosophy on concepts such as ``influence'' \cite{sunstein2016}, ``manipulation'' \cite{noggle2022}, ``mental interference'' \cite{douglas2021}, ``nudging'' \cite{schmidt2020}, ``choice architecture'' \cite{selinger2011}, ``persuasion'' \cite{berdichevsky1999}, ``cognitive liberty'' \cite{sententia2004}, ``mental integrity'' \cite{douglas2021}, ``mental self-determination''~\cite{bublitz2014}, freedom of thought \cite{lavazza2018}, and preference change \cite{carroll2021}. The definition of each of these terms, and the situations in which the phenomena they describe can be considered ethical, are all contested. We do not attempt to stipulate definitions or resolve normative disagreements in this paper. Rather, we draw on this literature to highlight specific reasons why some types of influence might be unethical, and link these concerns to relevant work from computer science and artificial intelligence (AI). Based on this synthesis, we propose a framework intended to help designers of algorithmic systems that influence people think more concretely about the ethical considerations relevant to their work.

    Before introducing the framework we would like to stress that not all influence is bad or morally questionable. Our definition of influence (given in the opening sentence of this article) is so broad as to encompass all causal relationships between agents. In this view, all human communication---much of which is beneficial---constitutes a form of influence. In particular, rational persuasion (``the unforced force of the better argument'' \cite{habermas1996}) is often delineated as being a morally acceptable form of communication, and hence influence \cite{SpahnLeadUsNot2012}. Without aiming to provide a perfect characterisation of wrongful influence, our view is that influence is ethically acceptable unless it possesses a property which makes it wrongful, and this paper is an attempt to compile a list of such properties.
    
    \paragraph{Method.} To arrive at the SHAPE framework we conducted an extensive (but unstructured) literature search in order to compile a list of of reasons why influence may be unethical. This longlist of reasons was iteratively grouped into sets of similar concerns, discussed, supplemented with additional literature searches, and re-grouped until we arrived at the current version of the framework. This process was not straightforward due to the fact that a number of categories have a non-negligible overlap, and our decisions about how to hierarchically arrange the relevant ideas are inevitably somewhat contingent and subjective. Additionally, we emphasise that this article is not a true systematic review, and the amount of literature relevant to the ethics of influence is vast. Nonetheless, we are confident that the chosen categories are informative, if not perfectly disjoint.

\section{Reasons why influence may be unethical}
\label{sec:concerns}

    In this section we develop our SHAPE (Secrecy, Harm, Agency, Privacy, Exogeneity) framework by considering reasons why influence may be unethical, drawing on work from moral philosophy and linking it to relevant concepts in computer science and AI. We do not claim that this set of reasons is comprehensive, but we do think it covers the most commonly-cited objections to influence. Similarly, we do not claim that this is a perfect taxonomy or that each of the reasons given is perfectly distinct from the others, but we do argue that the five groups of reasons---secrecy, harm, agency, privacy, and exogeneity---capture meaningful families of objections. An overview of the framework summarising these reasons is given in Box \ref{box:summary}.

    \begin{mybox}[t]
        \setlength{\fboxsep}{5mm}
        \begin{center}\noindent\fbox{\begin{minipage}{110mm}
            \sffamily

            \begin{center}\large The SHAPE framework\end{center}
            \begin{center}Influence might be unethical if...\end{center}
            \vspace{-2mm}
            {\color{newgrey}\hrule}
            \vspace{3mm}
            \textbf{\underline{S}ecrecy} (\ref{sec:secrecy})
            
            \begin{itemize}
                \item[...] the influencee is not fully aware of the intent to influence
                \item[...] the influencee is not fully aware of the means of influence
            \end{itemize}
            
            \textbf{\underline{H}arm} (\ref{sec:harm})
            
            \begin{itemize}
                \item[...] it causes harm
            \end{itemize}
            
            \textbf{\underline{A}gency} (\ref{sec:agency})
            
            \begin{itemize}
                \item[...] it removes options
                \item[...] it imposes conditional costs or offers
                \item[...] it occurs without consent
                \item[...] it bypasses reason
                \item[...] it is not able to be resisted
            \end{itemize}
    
            \textbf{\underline{P}rivacy} (\ref{sec:privacy})
            
            \begin{itemize}
                \item[...] it breaches an assumed contract regarding the use of personal information
                \item[...] it is implicated in mass surveillance
            \end{itemize}

            \textbf{\underline{E}xogeneity} (\ref{sec:exogeneity})

            \begin{itemize}
                \item[...] it serves exogenous rather than endogenous interests
                \item[...] it gives one group power over another
            \end{itemize}

        \end{minipage}}\end{center}\vspace{-3mm}
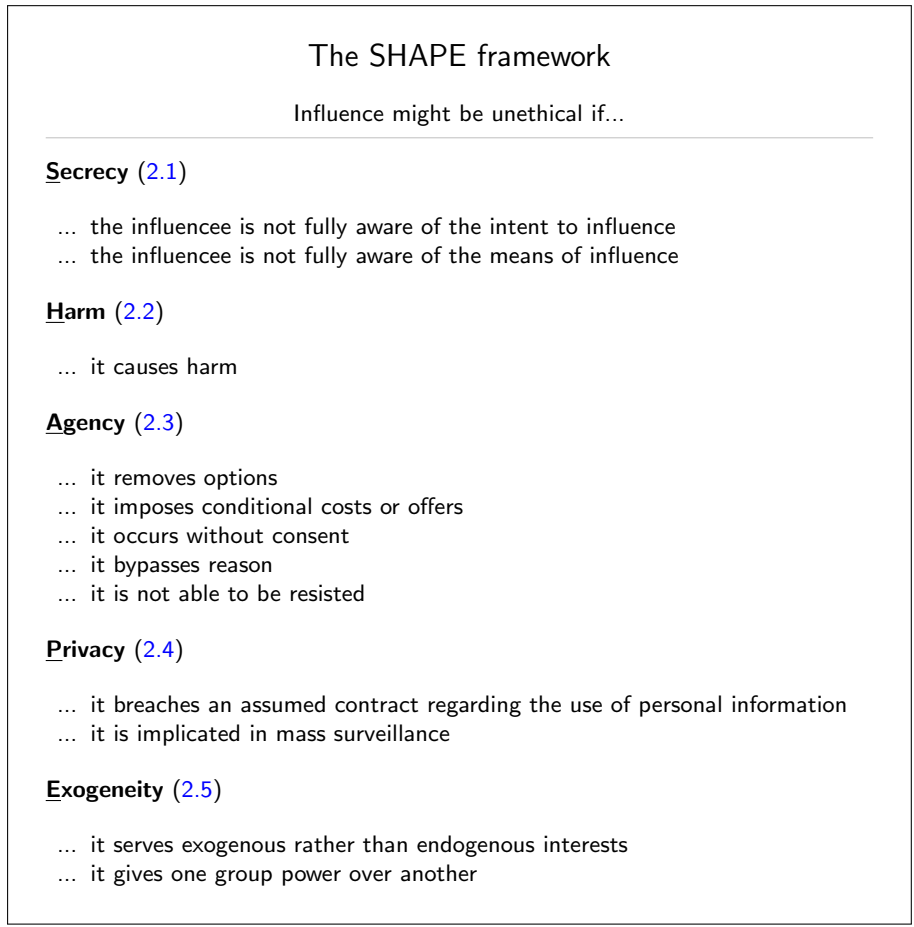
        \caption{The SHAPE framework for considering the ethicality of influence.}
        \label{box:summary}
    \end{mybox}

\subsection{Secrecy}
\label{sec:secrecy}

    First, influence may be unethical if it involves \textit{secrecy}. In the literature, variations of this idea have also been referred to as ``covertness'' \cite{dynel2016comparing}, ``deception''~\cite{martin2009philosophy}, ``lying'' \cite{mahon2018}, or ``trickery'' \cite{noggle2022}. Articulating precise definitions for these terms is an open philosophical problem \cite{levine2014encyclopedia}, but many have been proposed. The core idea is perhaps most neutrally defined as an ``information asymmetry'', where the influencer has more information than the influencee \cite{dierkens1991information}. More narrowly, deception has been defined as any situation where an agent \textit{A} intentionally causes another agent \textit{B} to have a false belief, with necessary requirement that agent \textit{A} does not believe it to be true \cite{carson2010lying}.

    Secrecy of all sorts may be wrong---when it is wrong---because it violates a moral norm or duty, specifically ``a duty to take care not to cause another to form false beliefs based on one's behaviour, communication, or omission'' \cite{shiffrin2014}, because it constitutes a breach of an implicit promise to be open and truthful \cite{ross2011}, or because it constitutes a betrayal of trust \cite{mahon2018}. The wrongness of secrecy may also in some cases be due to downstream consequences of the secrecy, rather than due to the secrecy itself. For example, some argue that when an intent to influence is hidden from the influencee, it is ``less likely to trigger rational scrutiny'' \cite{noggle2022} and thus bypasses reason, reducing agency (Section \ref{sec:agency}).

    That said, secrecy may not always be unethical, as in cases of ``benevolent deception'' \cite{adar2013benevolent}. For example, it may be beneficial for the rehabilitation of patients who have suffered stokes or other brain injuries if their physical therapist robot obfuscates their true progress towards recovery \cite{brewer2005perceptual}.
    
    Here, we distinguish between two types of secrecy as it relates influence: secrecy of \textit{intent} and secrecy of \textit{means}.
    
\paragraph{Secrecy of Intent.}

    Influence may be unethical if the influence is intended by the influencer, and the influencee is not fully aware of this intent. For example, a video deepfake \cite{mirsky2021creation} intended to influence public opinion in a certain direction (perhaps by misrepresenting the actions of a political figure) may be unethical because the people who are influenced are not made fully aware of this intention. Had they been aware, they would have assigned less credence to the information contained in the video \cite{lewandowsky2021}. See Box \ref{box:intentionality} for a more general discussion of intent as it relates to the ethics of influence.
    
\paragraph{Secrecy of Means.}

    Influence may also be unethical if the influencee is not fully aware of the means by which they are being influenced. For example, a user interacting with a sophisticated social media recommender system may be fully aware that the algorithm is designed to maximise the total amount of time they spend on the platform---so there is no secrecy of intent---but be unaware of the strategies the recommender is employing to achieve this, such as through the occasional recommendation of content that is increasingly sympathetic to a conspiracy theory \cite{thorburn2022}.

\subsubsection{Technical Work}

    Of the many ethical objections to influence, secrecy has perhaps received the most attention in the context of AI. For example, the sizable literature on algorithmic transparency, explainability, and interpretability (see, e.g., \cite{linardatos2021,carvalho2019}) represents an attempt to mitigate information asymmetries between AI systems and their human users. There is also an emerging literature that seeks to provide formal definitions of deception from a causal perspective, along with mechanisms for detecting it in AI systems \cite{sahbane2023,ward2023,ward2023a,ward2022}. Algorithmic agents can also fall prey to influence involving secrecy, as in cases of adversarial attacks \cite{madry2017towards}, data-poisoning \cite{madry2017towards}, reward function tampering \cite{everitt2021reward}, and manipulating human feedback \cite{ward2022agent}.

\newpage

\subsection{Harm}
\label{sec:harm}

    Second, influence may be unethical if it causes \textit{harm}. There are many different forms of harm, with some of the most prominent categories including reduced physical or mental well-being \cite{WHOEthicsGovernanceArtificial2021}, bias \cite{WeidingerEthicalSocialRisks2021}, unfairness \cite{WeidingerEthicalSocialRisks2021}, or injustice \cite{SunsteinEthicsNudging2015}. In general, harm and related concepts such as ``suffering'' \cite{hofmann2017} are expansively but inconsistently defined. Definitions range from those that equate harm with any ``physical or other injury or damage'' \cite{cambridge_dictionary}, to those state harm is a condition of ``interference with individual liberty'', originating from the ``harm principle'' of John Stuart Mill \cite{peczenik1995}, a definition which would liken harm to a reduction in agency (Section \ref{sec:agency}).

    Ethical (if not legal) views on what does and doesn't count as harm are normative and contested, and this is notably true of harms that may arise from speech acts in algorithmically-mediated online fora. For example, ``safe spaces'' are viewed by some as a means of avoiding psychological harm and others as an institution which, if realised, inflicts epistemic harms \cite{anderson2021}. Regardless of the position one takes in such debates, it seems defensible that there are many forms of harm which are widespread but not frequently well-articulated, and some of these harms can plausibly be promulgated by influential AI systems. One example of such harms has been labelled epistemic injustice \cite{KiddRoutledgeHandbookEpistemic2017}. Varieties of epistemic injustice include \textit{testimonial injustice}, where an individual is discredited as a credible source of knowledge, and \textit{hermeneutic injustice}, where an individual experiences reduced capacity to make sense of their own experiences due to a lack of a relevant framework, shared vocabulary, or common knowledge of a shared experience. Both forms of epistemic injustice may be exacerbated by language models or recommender systems, if such systems are heavily used and systematically privilege certain perspectives.
    
    It should be emphasised that harm, while perhaps intrinsically injurious, need not always be unethical. A surgeon making a cut to a patient's skin to fix their broken leg may cause temporary harm and pain, but is arguably acting in the best interests of the patient. In such cases, influence would then not be unethical despite causing harm. The assessment of harmful influence is further complicated by the fact that it can be very hard to define when influence is actually harmful, particularly influence over mental properties such as preferences \cite{carroll2021}.
    
\subsubsection{Technical Work}

    Given the breadth of the concept of harm, the goal of harm reduction can arguably be linked to almost all technical work on responsible or ethical AI. This is particularly true of work on the ``alignment problem'', which for many researchers is motivated at least in part by a utilitarian argument seeking harm reduction in the very long term \cite{gabriel2020}.
    
    Machine learning algorithms that seek to maximise reward or minimise regret map closely to a principle of harm reduction, though of course any given formally specified reward function may not capture harms that humans care about \cite{manheim2018}. The design of reward functions that can serve as valid measures of harm is in most domains a significant open problem. Close collaboration between technical researchers and those with domain specific expertise is a promising approach for designing formal metrics that capture qualitative, context-specific notions of harm \cite{lee2019}.
    
    \begin{mybox}[t]
        \setlength{\fboxsep}{5mm}
        \begin{center}\noindent\fbox{\begin{minipage}{110mm}
            \sffamily

            \begin{center}\large Intent\end{center}

            Whether influence is ethical may depend in part on whether it was intended, or whether a problematic property of that influence (e.g., secrecy) was intended. In law, whether intent (or more broadly, \textit{mens rea}) was present can determine the severity of the punishment or compensation. For example, intent is the difference between murder and manslaughter, and the difference between negligent and intentional infliction of emotional distress \cite{naffine2019}. In moral philosophy, the ``doctrine of double effect'' is the claim that it is sometimes permissible to cause harm as an ``unintended and merely foreseen side effect'' of doing good, even though it would not be permissible to cause such harm as a means of doing that same good \cite{sep-double-effect}. Views on the general relevance of intent to ethics differ \cite{shaw2006}. 
            
            \vspace{2mm}
            There may also be an ethically relevant distinction between human intention, and the intention of algorithmic agents (to the extent that such agents can be said to have intentions). The current majority view among machine ethicists and policymakers seems to be that while an autonomous algorithmic system may make ethical decisions, it is the human designers or deployers of those systems who bear moral responsibility for the actions of the system \cite{anderson2011,BathaeeArtificialIntelligenceBlack2017}. That said, various tests of intent commonly used in legal settings may break down when applied to cases involving inherently unpredictable or or incomprehensible black box algorithms~\cite{BathaeeArtificialIntelligenceBlack2017}.
            
            \vspace{2mm}
            \textbf{Technical Work}
                There are a few attempts to define intention in ways that can be applied to algorithmic systems, mostly by drawing analogies to how human intent is judged in legal processes \cite{AshtonDefinitionsIntentSuitable2023}. In one account, an algorithmic system is said to intend to perform a behaviour ``if, in performing the behaviour, the system can be understood as engaging in a reasoning or planning process for how the behaviour impacts some objective'' \cite{carroll2023}. More formal definitions of intent have also been proposed that define intent as a function of epistemic states, actions, and possible outcomes \cite{halpern2018}.
            
        \end{minipage}}\end{center}\vspace{-3mm}
        \caption{The concept of \textit{intent}, as it relates to the ethics of influence.}
        \label{box:intentionality}
    \end{mybox}

\subsection{Agency}
\label{sec:agency}

    Third, influence may be unethical if it reduces human \textit{agency}, or related concepts such as ``self-determination'' \cite{bublitz2014} and ``autonomy'' \cite{RubelAutonomyAgencyResponsibility2021}. There are many proposed definitions of agency \cite{FerreroIntroductionPhilosophyAgency2022}. One account defines agency as the act of an agent making use of its ability to act \cite{SchlosserAgency2019}. In this view, agency requires that executed actions are intended, and result in part from the agent's reasoning processes. To reduce human agency, then, is to disrupt the link between an agent's intentions or reasoning processes and their subsequent actions.

    Several works link influence with a reduction in agency. Being influenced into performing an action reduces the agency of an individual, at least in terms of the decision about whether to perform that action \cite{TaylorPracticalAutonomyBioethics2009}. Human agency is often characterised as having intrinsic moral value, and reductions in agency may be wrong regardless of whether that reduction in agency is paternalistic and results in improved welfare for the person affected. Not respecting the competency of an individual to make their own decisions is seen as a lack of appreciation of them being a rational agent \cite{SeymourFahmyLoveRespectInterfering2011} or even a degradation \cite{NoggleManipulativeActionsConceptual1996}. Perhaps more unambiguously, reduced agency can be wrong if it involves impairments to the psychological capabilities of the subject thought to be the basis for free will \cite{SripadaWhatMakesManipulated2012}. The wrongness of reductions in human agency may also stem from the fact that the interests of the affected agent are being devalued or deprioritised relative to those of the another party (see Section \ref{sec:exogeneity}) \cite{RudinowManipulation1978,SeymourFahmyLoveRespectInterfering2011}.
    
    However, it has also been argued that reductions in agency are not always wrong, and that rational agents often do not oppose influence that has this effect \cite{BussValuingAutonomyRespecting2005}. Instead, agency may be valuable instrumentally because is often a useful means to an end. We sometimes place ourselves in situations where we have reduced agency---such as following a recipe or studying a prescribed curriculum---if it helps to achieve a goal.

    Here, we give five accounts of what it means for influence to reduce agency: removing options, imposing conditional costs or offers, influencing without consent, bypassing reason, or being irresistible. These are likely not mutually exclusive.

\paragraph{Removal of Options.}

    Influence may be unethical if it removes options previously available to the influencee \cite{garnett2017}. For example, an autonomous vehicle may in some implementations prevent its human driver from deciding to take a certain route to a destination that they otherwise would have taken. Options may be removed explicitly (by refusal) or implicitly (by a failure to provide an affordance that would enable the option). Options can also be removed effectively, without being absolutely removed, by imposing conditional costs (see below) that are so severe as to make the option untenable. Such removal of options, where the influenced party can be said to have no choice or no acceptable choice, has been labelled ``coercion'' \cite{nozick1969,kligman1992,wood2014}.

\paragraph{Conditional Costs or Offers.}

    Influence may be unethical if it imposes conditional costs or offers on the influencee depending on the action they choose to take, thus altering the relative appeal of different options. In philosophical literature, this type of influence is sometimes called ``pressure'' \cite{noggle2022}. Conditional costs can be seen as a form of threat, though the severity of the threatened cost can vary significantly. Examples of costs that might be threatened include a loss of time or energy (e.g., nudging \cite{SunsteinEthicsNudging2015} or browbeating \cite{baron2003}), a loss of social status (e.g., peer pressure), or physical violence (e.g., kidnappers demanding a ransom). 

    It is possible to use carrots as well as sticks---the costs imposed may be opportunity costs. For example, the influencer may attach positive incentives or ``offers'' (e.g., money or status) to certain alternatives, which reduces the relative value of others \cite{sachs2013}. Such incentives are not always unethical. For example, it is generally considered acceptable to offer salaries to influence people to work for you. One account \cite{baron2003} suggests that such incentives are only unethical if they mean the influenced adopts a particular alternative for ``the wrong sort of reason'' \cite{baron2003}. Which sorts of reasons are considered wrong will be context specific.

\paragraph{Consent.}
\label{ssec:consent}

    Influence may be unethical if it occurs without (informed) consent, thus potentially ignoring a decision a person has made while exercising their agency \cite{faden1986}. Consent ``[renders] permissible otherwise impermissible actions'' \cite{jones2018}, and thus may in some contexts be a necessary condition for ethical influence. For example, consent is plausibly the morally distinguishing factor between strenuous exercise and forced labour, or to give a more algorithmic example, between ethical and unethical online targeted advertising. It may be particularly challenging to obtain informed consent in the context of AI systems, which may not be comprehensible to lay users, and to which there may be no alternatives of comparable capability \cite{andreotta2022}.

\paragraph{Bypassing Reason.}

    Influence may be unethical if it bypasses human reason \cite{GorinManipulatorsAlwaysThreaten2014}. Mechanisms of influence which involve the bypassing of reason include: customised presentation of information, the flooding of agents with irrelevant information to crowd out relevant information, and the withholding of certain information \cite{Blumenthal-BarbyFrameworkAssessingMoral2014}; exploitation of known imperfections in human decision-making such as group pressure \cite{AschOpinionsSocialPressure1955}; exploitation of the ``truth effect'', which is the fact that frequent repetition of a statement increases the probability of individuals to find that statement to be true \cite{HasherFrequencyConferenceReferential1977,SchwartzRepetitionRatedTruth1982}; anchoring \cite{AdomaviciusRecommenderSystemsManipulate2013}; and appeals to emotion such as fear \cite{HowardIRASocialMedia2019}.

\paragraph{Irresistibility.}

    Influence may be unethical to the extent that it is difficult to resist \cite{Blumenthal-BarbyFrameworkAssessingMoral2014,CaveWhatWrongMotive2007}. Attempts at influence can be made difficult to resist through the use of techniques such as flattery or seduction. Use of such techniques arguably reduces agency of those influenced. This has direct implications on the moral responsibility of an agent for their actions. Such responsibility has been claimed to not require ``regulative control'', i.e. access to alternative possibilities, but merely ``guidance control'' as control over the mechanism which steers their behaviour. An agent who is influenced into acting in a certain way through mechanisms they cannot resist is therefore not morally responsible for the consequences of their actions \cite{FischerResponsibilityManipulation2004}.

\subsubsection{Technical Work}

    There is an emerging body of technical work that seeks to quantify degrees of agency, often from a causal perspective \cite{kenton2022,chan2023}. There has also been work that seeks to use AI to support human agency in certain contexts, such as in learning environments \cite{DeschenesRecommenderSystemsSupport2020} or on social media platforms \cite{KangAIAgencyVs2022}.

\subsection{Privacy}
\label{sec:privacy}

    Fourth, influence may also be unethical if it is made possible by a violation of \textit{privacy}, or related constructs such as the ``contextual integrity'' of information flows \cite{nissenbaum2004privacy}. The more information is known about a person, the greater the extent to which it is possible to identify mechanisms by which they can be influenced, such as via the personalisation or microtargeting of persuasive messages \cite{tappin2023}.
    
    Justifications for privacy (or the right to privacy) vary \cite{solove2008}. One prominent account articulates three objectives are frequently cited when justifying privacy-enhancing laws: (1) limiting surveillance of citizens and use of information about them by agents of government, (2) restricting access to sensitive, personal, or private information, and (3) curtailing intrusions into places deemed private or personal~\cite{nissenbaum2004privacy}.

    In the context of AI, different privacy considerations may apply to different kinds of data. One account distinguishes between the following kinds: \textit{training} data (i.e., data collected to train predictive models) vs \textit{targeting} data (i.e., data used for targeting); \textit{sensitive} data (i.e., data about a person that they might reasonably not want others to know) vs \textit{nonsensitive} data. Sensitive data can besubdivided further into \textit{intrinsically sensitive} data (i.e., if it is sensitive when considered on its own) vs \textit{extrinsically sensitive} data (if it is sensitive only when considered in combination with other data points) \cite{BennWhatWrongAutomated2022}. Privacy concerns arise when the training data consists of sensitive and nonsensitive information \cite{barocas2014big}; a model trained on that data can uncover a link between intrinsically nonsensitive properties $P$, $Q$, and $R$, and intrinsically sensitive property $S$. This means that if we have access to values for these non-sensitive properties for a user, the chances of successfully predicting $S$ increase~\cite{BennWhatWrongAutomated2022}.
    
    Here, we delineate two kinds of privacy violation that can enable influence: individual breaches of contract, and mass surveillance.
    
\paragraph{Breach of Contract.}

    Influence may be unethical if it is enabled by the breaching of a contract (explicit or implicit) that was understood by the influencee to constrain access to or use of personal information. For example, a recommender system might learn user embeddings that contain inferred, sensitive information such as sexuality, political leaning, or degree of mental health, even though such information was not consciously contributed by the user and not understood by them to be inferred \cite{seaver2021}. This information might then be used to tailor content recommended to the user, breaching this assumed contract. Plausibly, such a privacy violation could also be framed as a lack of consent (Section \ref{sec:agency}).
    
\paragraph{Mass Surveillance.}

    Influence may also be unethical if it is implicated in mass surveillance. In this view, it is not (only) the access to or use of personal information that is wrongful, but the initial collection of that data. Mass surveillance or data collection, due to its scale, increases several risks, including but not limited to secrecy about the means of influence (Section \ref{sec:secrecy}), and the capability to reduce agency (Section \ref{sec:agency}). This gives the surveillant---be it government, a corporation, or other entity---power over the surveilled (Section \ref{sec:exogeneity}). Given the combination of these concerns, mass surveillance is often treated as intrinsically wrongful \cite{Zuboff}.

    While mass surveillance is in some ways merely the aggregation of many individual breaches of contract, it is often discussed separately and defended against using separate mechanisms, so we discuss it separately here. We note also that in the context of AI, the data of individuals is often only useful for model training when used in aggregate with the data of many others, and thus privacy violations that facilitate influential AI systems are likely to occur ``at scale''.

\subsubsection{Technical Work}

    While the most obvious approaches to mitigating privacy concerns relating to influence involve simply deciding whether or not to proceed with a given product deployment or research project, there is also research on technical approaches to respecting privacy in certain applications of influential AI. These include work on differential privacy \cite{dwork2006,abadi2016} and contextual integrity \cite{benthall2017,criado2015}.

\subsection{Exogeneity}
\label{sec:exogeneity}

    Lastly, influence may be unethical if it advances interests not held by the agent being influenced, a property we call \textit{exogeneity}. We present two articulations of unethical exogeneity in influence: the disparate advancing of exogenous and endogenous interests, and the exercise of power.

\paragraph{Exogenous Interests.}

    Influence may be unethical if it advances exogenous goals or interests (those not held by the influencee) over endogenous goals or interests (those held by the influencee). In this account, the wrongness of influence stems not from the fact that the influencer benefits (they may not benefit), or from harm to the influencee in absolute terms (they may not be harmed), but from the relative advantaging of the interests of another agent over the interests of the influencee \cite{BennWhatWrongAutomated2022,noggle2022,RudinowManipulation1978}.
    
    An example of this would be the use of recommender systems to manipulate elections. Such systems are frequently used by social media sites and were shown to have significant effects on the voting behaviour of individuals due to them changing the political content that users see \cite{Ndlela2020}. If a recommender system is manipulated by a third party in order to bias users towards voting for a specific political candidate which does not represent their interests, this would constitute an example of unethical influence due to that third party preferring its interests over those of the user.

\paragraph{Power.}

    Influence may also be unethical if it empowers one party over another, or constitutes an exercise of power of one party over another. There is considerable philosophical literature on how power is instantiated in technology \cite{bloomfield1992information}, as well as related concepts including ``control'' and ``domination'' \cite{aytac2022}. For example, manipulating the opinion of a single individual can be difficult \cite{coppock2020}, but widely-used recommender systems present a vector by which a minority might steer the opinions and behaviour of a larger population, through an accumulation of small or stochastic effects.

    Power does not need to be acted upon to be unethical. One account states that freedom from domination requires not just that one is ``let alone'', but that one is free from the possibility of ``reserve control'' that could be enforced any time at the discretion of the dominating party \cite{pettit2014}. It is also important to recognise that there are also several mechanisms via which the design of technology, including AI systems, can influence the moral beliefs and practices of a population, a phenomenon referred to as technologically-mediated moral change \cite{danaher2023}. In this way, arguably all designers of all algorithmic systems have power over the societies into which they deploy their creations. 
    
\subsubsection{Technical Work}

    Monitoring whose interests are being served through the use of an AI system lends itself naturally to questions of fairness, and there is substantial literature on both formal measures of fairness \cite{fairnessmetricssurvey} and algorithms for promoting it \cite{biasmitigationmethodssurvey}. Another relevant line of work relates the development of mechanisms for diffusing or decentralising the power that is exercised through the use of influential AI systems. This includes both technical social choice mechanisms for choosing objective functions \cite{lee2019}, and the use of participatory institutions such as citizen assemblies \cite{ovadya2021} and collective response systems \cite{ovadya2023,cip2023} to provide democratic oversight. Finally, there is early work on how the influential power of an online platform might be quantified and monitored \cite{hardt2022}.

\section{Governance of Influence}

    For the most part, the concerns listed in Section \ref{sec:concerns} point to general or abstract principles that can inform an understanding of the ethical status of different kinds of influence. In order for such an understanding to be widely adopted into the practices of those designing and building influential algorithmic systems, we need mechanisms for deciding, disseminating and enforcing what best practice looks like in specific, concrete terms. Here we point to three such mechanisms (professional cultures, ethics review processes, government regulation) via examples from other domains (scientific research with human subjects, journalism, advertising).

\subsection{Professional Culture} \label{ssec:journalistic_ethics}
    
    In journalism there is minimal formal oversight of ethical practice, but nonetheless there is broad understanding of a core set of ethical principles which are reinforced by educational institutions, professional organisations, and workplace culture \cite{frost2015,sanders2003}. These principles commonly include mention of accuracy or truthfulness \cite{porlezza2019}, objectivity or impartiality \cite{ward2019}, and avoidance of harm through the use of anonymity or avoiding coverage of certain topics (e.g. suicide) \cite{carlson2010,domaradzki2021}. Such principles informally govern influence in the context of journalism.
    
    Similar ethical principles exist in computer science, but these are not as widely adopted \cite{bcscodeofconduct,acmcodeofconduct}.

\subsection{Institutional Ethics Reviews} \label{ssec:ethics_reviews}
    
    Formal ethics review processes, such as those conducted by most academic institutions in advance of research that involves human subjects, are one way of formalising a consideration for the ethics of influence. Reviewers involved in such processes already grapple with the use of techniques such as deception or trickery to create experimental conditions \cite{athanassoulis2009}, and with what it means to have meaningfully consented to be subject to such influence \cite{hoeyer2014}.
    
    Ethics review processes are already used by conferences and journals in computer science \cite{SrikumarAdvancingEthicsReview2022}, as well as by companies that build and deploy AI systems \cite{AdobeAIEthics,GoogleAIReview}, though they often lack a substantive focus on influence.
    
\subsection{Regulation} \label{ssec:advertising_regulation}
    
    In many jurisdictions, the advertising industry is subject to laws that place limits on the content of advertising and the contexts in which certain types of advertising can appear. These often require that advertising avoid outright deception (e.g., truth-in-advertising laws) \cite{tushnet2008}, and ban ads in contexts where they are thought to cause harm (e.g., the ban of gambling, alcohol, or fast food ads during childrens' programs or televised sports) \cite{adams2012,thomas2018}. Such laws formally specify classes of influence which are collectively deemed unacceptable in the context of advertising.

    Several proposed or enacted regulations focused on AI and algorithmic systems are relevant to aspects of influence. Many jurisdictions have regulations focused on privacy, the most prominent of which is perhaps the European Union's (EU) General Data Protection Regulation (GDPR) \cite{ParliamentImpactGeneralData2020}. If approved in its current form, the proposed EU AI Act \cite{EUAIAct2023} would prohibit certain forms of influential AI. Specifically, Article 5 of the current draft text states
    \begin{quotation}
        \noindent 
        The following artificial intelligence practices shall be prohibited:
        \begin{itemize}
            \item[(a)] the placing on the market, putting into service or use of an AI system that deploys subliminal techniques beyond a person’s consciousness or purposefully manipulative or deceptive techniques, with the objective to or the effect of materially distorting a person’s or a group of persons behaviour by appreciably impairing the person’s ability to make an informed decision, thereby causing the person to take a decision they would not have taken otherwise in a manner that causes or is likely to cause that person, another person or group of persons significant harm; ...
        \end{itemize}
    \end{quotation}
    In subsequent clauses, the draft act also prohibits other forms of influential AI~\cite{EUAIAct2023}.

\section{Conclusion}

    In this paper we have synthesised some of the most commonly cited reasons---captured by the acronym SHAPE---why influence can be unethical. Specifically, these are that influence can (1) involve \textit{secrecy} regarding the intent or means of influence, (2) cause \textit{harm}, (3) reduce human \textit{agency} by removing options, imposing conditional costs or offers, occurring without consent, bypassing reason, or being irresistible, (4) violate \textit{privacy} by relying on the use of private information in a way that breaches an assumed contract or being implicated in mass surveillance, and (5) advance \textit{exogenous} interests at the expense of endogenous interests, or give one group power over another. We linked each of these general principles to relevant concepts from computer science and artificial intelligence, and described three models of ethical governance from other domains---professional culture which emphasises ethics, institutional ethics reviews, and regulation---which could be employed to translate such general principles into practice.

    We envisage the SHAPE framework being used by designers of influential AI systems as a way to structure their thinking when considering the ethical impacts of their systems. For example, those building a product based on a large language model (LLM) might systematically work through Box \ref{box:summary}, enumerating the examples of each of the SHAPE concerns that arise in the context of their product. These might include user-to-LLM feedback loops that are not understood by the user (\textit{secrecy}), defamatory hallucinations (\textit{harm}), affordances that require extra effort by users to surface certain perspectives in model outputs (\textit{agency}), use of personal data to improve user retention (\textit{privacy}), and adversely paternalistic choices in the design of the product (\textit{exogeneity}), among others. Such a list could then be translated into a list of actions to be taken to remove or mitigate each of these ethical concerns.

    For the most part, we have in this paper refrained from stipulating particular definitions or drawing definitive lines between ethical and unethical influence. Such decisions will likely be context-specific and contested, and our focus has instead been on drawing connections between work in philosophy and computer science. That said, it would be valuable for future work to consider the extent to which these concerns over influence could be made more precise by focusing on narrower domains, such as LLM-enabled chat interfaces or social media recommender systems.

\subsubsection{Acknowledgements.} The authors were supported by UK Research and Innovation [grant number EP/S023356/1], in the UKRI Centre for Doctoral Training in Safe and Trusted Artificial Intelligence (\href{https://safeanadtrustedai.org}{safeandtrustedai.org}), co-located at King's College London and Imperial College London.

\bibliographystyle{splncs04}
\bibliography{references}

\end{document}